\documentclass[journal=jacsat,manuscript=article,nomcite]{achemso}

\SectionNumbersOn
\usepackage[version=3]{mhchem} 
\usepackage{natbib}
\setkeys{acs}{maxauthors=0}  
\usepackage{pdfpages}
\usepackage{xr-hyper}
\usepackage{hyperref}
\usepackage{xspace}           
\usepackage{xcolor}
\usepackage{geometry}
\usepackage{graphicx}
\usepackage{amsmath}
\usepackage{amsfonts}
\usepackage{subfigure}
\usepackage[most]{tcolorbox}
\usepackage{booktabs}
\usepackage{makecell}
\usepackage{ulem}
\usepackage[ruled,vlined]{algorithm2e}
\usepackage{xcolor}

\definecolor{selfgreen}{RGB}{200,235,200}
\definecolor{backred}{RGB}{240,200,200}

\newtcolorbox{SelfRecBox}{
  enhanced,
  colback=selfgreen,
  colframe=selfgreen,
  boxrule=0pt,
  left=2mm,
  right=1mm,
  top=1mm,
  bottom=1mm,
  title=\textcolor{green!60!black}{\textbf{Self-recurrence loop}},
  coltitle=green!60!black,
  fonttitle=\bfseries,
  before skip=2mm,
  after skip=2mm
}

\newtcolorbox{BackGuidBox}{
  enhanced,
  width=0.95\linewidth,
  colback=backred,
  colframe=backred,
  boxrule=0pt,
  left=2mm,
  right=1mm,
  top=1mm,
  bottom=1mm,
  title=\textcolor{red!70!black}{\textbf{Gradient descent loop}},
  coltitle=red!70!black,
  fonttitle=\bfseries
}

\newtcolorbox{CorrectorBox}{
  enhanced,
  width=0.95\linewidth,
  colback=blue!30!white,
  colframe=blue!30!white,
  boxrule=0pt,
  left=2mm,
  right=1mm,
  top=1mm,
  bottom=1mm,
  title=\textcolor{blue!70!black}{\textbf{Corrector loop}},
  coltitle=blue!70!black,
  fonttitle=\bfseries
}

\DontPrintSemicolon
\SetAlgoLined
\SetAlgoNoEnd

\newcommand{\N}{\mathcal{N}}

\author{Auguste de Lambilly} 
\affiliation{CNRS-Saint-Gobain-NIMS, IRL 3629, Laboratory for Innovative Key Materials and Structures (LINK), 1-1 Namiki, 305-0044 Tsukuba, Japan}
\alsoaffiliation{LTCI, Télécom Paris, Institut Polytechnique de Paris, France}
\author{Vladimir Baturin}
\affiliation{CNRS-Saint-Gobain-NIMS, IRL 3629, Laboratory for Innovative Key Materials and Structures (LINK), 1-1 Namiki, 305-0044 Tsukuba, Japan}
\author{David Portehault}
\affiliation{Laboratoire de Chimie de la Matière Condensée de Paris (LCMCP), Sorbonne Université, CNRS, 4 place Jussieu, F-75005 Paris, France}
\author{Guillaume Lambard}
\affiliation{Data-driven Materials Design Group, Center for Basic Research on Materials
, National Institute for Materials Science, Ibaraki, Tsukuba, Namiki 1-1, 305-0044 Japan}
\alsoaffiliation{CNRS-Saint-Gobain-NIMS, IRL 3629, Laboratory for Innovative Key Materials and Structures (LINK), 1-1 Namiki, 305-0044 Tsukuba, Japan}
\email{LAMBARD.Guillaume@nims.go.jp}
\author{Nataliya Sokolovska} 
\affiliation{Laboratory of Computational, Quantitative, and Synthetic Biology (CQSB), Sorbonne Université, CNRS, 4 place Jussieu, F-75005 Paris, France}
\author{Florence d'Alché-Buc}
\affiliation{LTCI, Télécom Paris, Institut Polytechnique de Paris, France}
\author{Jean-Claude Crivello}
\affiliation{CNRS-Saint-Gobain-NIMS, IRL 3629, Laboratory for Innovative Key Materials and Structures (LINK), 1-1 Namiki, 305-0044 Tsukuba, Japan}
\email{jean-claude.crivello@cnrs.fr}

\title[]
  {Fine-tuning-Free Diffusion Model with Adaptive Constraint Guidance for Inorganic Crystal Structure Generation}
  
\begin{document}
\textbf{\center Auguste de Lambilly and Vladimir Baturin contributed equally to this work.}
\newpage
\begin{abstract}
Generative diffusion models have emerged as powerful tools for the discovery of inorganic crystal structures, yet steering their sampling process toward user-defined physical and chemical objectives remains challenging. 

We present a computational framework that integrates adaptive constraint guidance into a pre-trained crystal diffusion model, enabling the generation of candidate structures that satisfy targeted structural and chemical requirements without model retraining. 
The approach incorporates differentiable constraint functions directly during sampling, providing an interpretable mechanism for expert-driven exploration of the crystal structure space.
To assess the reliability of generated candidates, we introduce a multi-stage validation workflow combining descriptor-based analysis, duplicate removal, comparison with reference crystal databases, graph neural network energy prediction, and thermodynamic stability evaluation through convex-hull analysis. 

The framework is applied to several classes of inorganic compounds and to constraints involving atomic volume, local coordination environments, and near-neighbor structural motifs. 
Results demonstrate that adaptive guidance effectively redirects the sampling distribution toward structures exhibiting the desired characteristics while preserving chemical plausibility. 
Subsequent validation reveals which generated candidates remain viable after energetic and thermodynamic screening.
The proposed methodology provides a practical and transparent strategy for incorporating expert knowledge into crystal generative models and establishes a general computational framework for constrained materials discovery.
\end{abstract}

\vspace{0.5cm}

\newpage
\section{Introduction} 

Identifying and employing new functional materials has become critical to address pressing environmental (abundance, sustainability) and geopolitical challenges~\cite{brechet14}.
The discovery of new inorganic solids relies on both empirical and computational approaches. 
Empirical methods, which combine chemical design rules with trial-and-error experiments, remain time- and resource-intensive. 
In contrast, computational methods---particularly first-principles approaches such as density functional theory (DFT)---have enabled a deeper understanding of matter and, in some cases, prediction of atomic-scale ordering, albeit at high computational cost~\cite{becke_perspective_2014}.

During the past decade, data-driven science and machine learning (ML) have emerged as powerful paradigms to accelerate materials discovery, especially by guiding experiments~\cite{agrawal_perspective_2016,ward_2016,isayev_2017,butler_2018}. 
Generative models, including Generative Adversarial Networks (GANs)\cite{nouira_crystalgan_2019,kim_generative_2020} and more recently diffusion models\cite{DDPM,diffusion_survey}, have enabled the direct generation of novel structures rather than the screening of large candidate spaces~\cite{sultanov_2023}.
In parallel, Physics-Informed Neural Networks (PINNs) incorporate physical laws into the learning process, ensuring consistency with fundamental principles~\cite{raissi_physics_2017}. 
Meanwhile, Graph Neural Networks (GNNs) and Machine Learning Interatomic Potentials (MLIP) have become standard tools for efficient and accurate property prediction, leveraging the relational structure of crystalline materials without requiring expensive DFT calculations~\cite{riebesell_2025}.

Despite these promising approaches, a critical gap has emerged: 
at the moment, data alone cannot replace domain expertise and critical thinking in materials science. 
This became evident with large-scale models developed by leading technology companies~\cite{yang2023scalable,uma_2025} announcing impressive claims, such as Google DeepMind's GNoME predicting 2.2 million new crystal structures~\cite{merchant_2023}. 
More precise analyzes revealed fundamental flaws: many predicted structures were merely combinatorial constructs of already known compounds, without proper experimental verification~\cite{cheetham_2024}. 
Subsequent analyzes also raised concerns about the claims of 41 new inorganic compounds reported by the autonomous laboratory for the accelerated synthesis of novel materials (A-Lab), as several structures were found to be mischaracterized or already known in the literature~\cite{szymanski_2023,leeman_challenges_2024}. 
These issues highlight that numerical predictions, without chemical intuition and physical validation, can lead to physically implausible or trivial "discoveries".

The purpose of this paper is to integrate critical human guidance grounded in chemistry into diffusion models, 
building on our prior work on diffusion-based models\cite{sultanov_2023} and on state-of-the-art generative model MatterGen\cite{zeni2025generative}.
By introducing an additional module of chemical constraints (such as minimum interatomic distances or local chemical environments), we allow chemists, leveraging their physical intuition, to generate novel, functional crystallographic structures that (i) satisfy a pre-defined set of constraints and (ii) are (meta)stable, meaning their energy lies below a pre-defined threshold above the convex hull. 
Rather than generating millions of unverified candidates, our approach emphasizes quality over quantity, ensuring that the generated structures respect fundamental chemical principles and domain-specific knowledge from the outset.
We focus on constraints whose satisfiability can be estimated through differentiable functions, such as neural networks. 
Adding new constraints should then be straightforward and should not require additional fine-tuning. 

To this end, our approach relies on a foundation model: an unconstrained generative model for inorganic materials that has been pre-trained to generate stable and diverse crystal structures across the periodic table.
Here, we employ MatterGen~\cite{zeni2025generative}, a robust diffusion-based model for inorganic materials. 
MatterGen has been pre-trained to generate stable structures by focusing on compounds within 0.10 eV/atom of the convex hull, ensuring diversity in crystal structures and supporting flexible property conditioning.
While MatterGen currently handles unit cells with up to 20 atoms only, this limitation is outweighed in our setting by its optimization and demonstrated performance, making it a strong backbone for our approach.
Although geometric descriptors can be computed directly, fine-tuning requires sufficient training examples around each target. 
Several constraints considered here correspond to rare or absent regimes in the available data, including dense phases, unusual coordination environments, and combined motifs in an unexplored quaternary system. 
Training-free guidance allows such objectives to be explored without constructing a dedicated dataset for each constraint.

In this work, we do not introduce a new general-purpose guidance algorithm. Instead, we adapt the Universal Guidance framework to the periodic and multicomponent representation of inorganic crystals used by MatterGen. 
Differentiable objectives based on atomic volume, local coordination environments, and combinations of near-neighbor descriptors are incorporated into the sampling process without modifying or retraining the pretrained score model. 
This formulation allows structural objectives to be introduced and combined at inference time, while retaining the same chemical-system checkpoint.
The main contribution of this work is therefore an integrated framework for formulating, applying, and evaluating training-free crystallographic constraints. 
Its performance is assessed through five case studies ranging from unary to quaternary systems. 
We evaluate not only the enrichment of the requested descriptors in the raw generated distributions, but also the retention of candidate structures after symmetrization, deduplication, reference-database matching, energetic screening, structural inspection, and targeted first-principles validation. 
The purpose is not to establish training-free guidance as a replacement for conditional fine-tuning or conventional crystal-structure prediction. 
Rather, we investigate its ability to act as a flexible proposal mechanism for querying a pretrained crystal generator with user-defined structural hypotheses, and we identify both the regimes in which this control produces useful candidates and those in which descriptor enrichment does not translate into an improved final yield.
\section{Fine-tuning-free guidance in diffusion models}

\subsection{Context of the score-based methods}

Generative models can exploit an iterative denoising process (as in diffusion models), a latent-variable formulation trained via variational inference (as in variational autoencoders), or an explicit likelihood-based construction, such as normalizing flows and autoregressive models~\cite{tomczak2022deep}.
In this work, we use a diffusion model that demonstrated impressive performance in the generation of crystal structures, based on the MatterGen model~\cite{zeni2025generative}.

The unit cell is defined by a triplet $(X,A,L)$, where $X$ denotes the fractional atomic coordinates, $A$ the atomic species, and $L$ the lattice basis (see SI~1 for details).
We describe below the diffusion process used in the model.
If not explicitly stated otherwise, we use the notation $z_t$ for the sample $z$ drawn from the distribution at time $t$.

\begin{figure}
    \centering
    \includegraphics[width=0.8\linewidth]{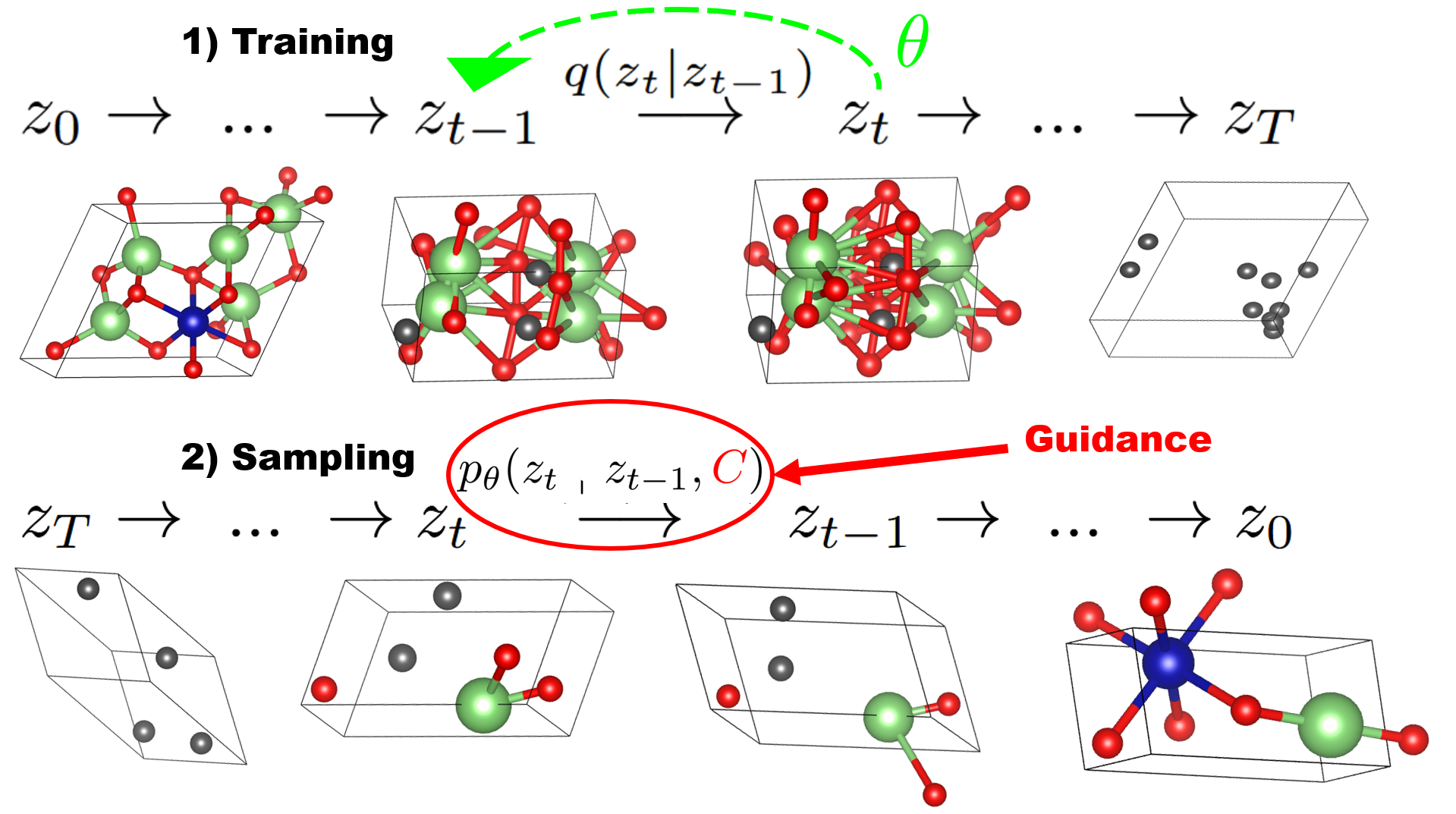}
    \caption{Diffusion model for material generation based on two processes: training and sampling. The green arrow represents parameters $\theta$ estimated in the training. The guidance step is shown in red, with condition $C$.
    }
    \label{fig:diffusion_model}
\end{figure}

Using the representation of a crystal structure described in SI~1, a diffusion model for compound generation is an algorithm that is trained to recover the observations in the backward process by denoising samples corrupted in the forward procedure.
The forward process -- denoted by $q$ -- is defined by $z_t|z_{t-1}\sim \N(a_tz_{t-1}, b_t^2 {\bf I})$, where $a_t$ scales the mean of the transition $z_t|z_{t-1}$, while $b_t$ controls the noise variance, jointly governing the gradual noising process in the forward diffusion, and $\bf I$ is identity.
The specific features of the forward noising process are described in SI~2.1. 
We assume that $z_t|z_0 \sim \N(\alpha_t z_0; \sigma_t^2 {\bf I})$, the complete proof can be found in SI~2.2. The relationship between $b$ and $\sigma^2$ is also detailed in the supplementary material.

We work in the context of score-based diffusion models. The score of the diffusion model used to generate samples is defined as 
\begin{equation*}
s_{\theta}(z_t) \approx \nabla_{z_t}\log q(z_t),
\end{equation*}
where $\theta$ are parameters learned during training. 
This score is directly involved in the denoising process: 
\begin{equation}
   z_{t-1}|z_t\sim\N\left(\dfrac{z_t+b_t^2s_{\theta}(z_t)}{a_t} ,\dfrac{{\sigma}_{t-1}^2}{{\sigma}_t^2}b_t^2\right).
    \label{eq:reverse}
\end{equation}

Figure~\ref{fig:diffusion_model} summarizes the diffusion process adapted to crystal structure generation. 
The green arrow represents parameter learning of the model. 
The model relies on the assumption that an observation cannot be distinguished from a Gaussian noise after $T$ noising steps. 
Once the model (the score) is trained, the generative process starts from a Gaussian noise and successively denoises a sample for a pre-defined number of timesteps. 

\subsection{Guidance in diffusion models}

The primary objective of generative models in materials science has traditionally been the generation of stable materials. 
However, in practice, researchers are often interested in specific materials that have to satisfy some well-defined properties and be experimentally reachable, even if they do not correspond to the most stable structures. 
For example, the synthesis of functional materials often involves strict constraints: the coordination number of metal centers can dictate optical properties, interatomic distances must be finely tuned to enable some catalytic mechanisms or provide mechanical properties, and the band gap must be adjusted to meet the requirements of optoelectronic applications.
Incorporating mechanisms to control the generation process within such constraints is, therefore, essential to make these models more suitable to address desired properties.
In the present work, we use guidance for this purpose.
\smallskip

Guidance methods for diffusion models~\cite{ho2021classifier,bansal2023universal,training-free-guidance} aim to modify the reverse diffusion process so that a certain property $C$ is satisfied at sampling time.
From a Bayesian point of view, the score function is modified to incorporate the condition $C$, yielding:
$$\nabla_{z_t} \log q(z_t,C)= \nabla_{z_t}\log q(z_t)+ \nabla_{z_t}\log q(C|z_t). $$

The score is approximated by the score network $\nabla_{z_t}\log q(z_t)$, and an estimate of $\nabla_{z_t}\log q(C|z_t)$ must be approximated. Some details on guidance for diffusion models and methods to compute the approximation are provided in SI~3.

\subsubsection{Universal Guidance}
MatterGen uses classifier-free guidance (CFG) trained on a set of selected stable compounds. 
The CFG allows to guide the backward diffusion process toward a desired result, such as a stable structure, without requiring an additional classifier layer (more details in SI~3.2).
The main issue with the CFG is that one needs to retrain the model for each specific new set of conditions. 
Otherwise, the model is out-of-distribution, and can yield irrelevant results, especially if the conditions are far from those used for training. 
To overcome this issue, the Universal Guidance method \cite{bansal2023universal} (details provided in SI~2) uses an off-the-shelf loss function as guidance for diffusion. 
We consider the following formulation: 
$$ \nabla_{z_t}\log q(C|z_t) \approx -\nabla_{z_t} \ell(C,f(z_t)). $$

Indeed, the negative log-likelihood is a measure of the model's performance
widely used in machine learning.
We use here the cross-entropy loss $\ell$ that measures how strongly the condition $C$ is violated by $z_t$;
$f$ is an objective function that evaluates a specific property of $z_t$ and compares it to condition $C$. 
However, estimating the property is challenging for noisy samples, especially if $f$ is a black-box machine learning 
model trained on clean samples only. 
The Universal Guidance method replaces $z_t$ by $\hat{z}_{0|t}$, the estimate of $z_0$ at timestep $t$, which is the predicted final sample:
$$ \nabla_{z_t}\log q(C|z_t) \approx -\nabla_{z_t} \ell(C,f(\hat{z}_{0|t})). $$

We recall that ${\sigma}_0=0$: $ \hat{z}_{0|t} = \dfrac{z_t+{\sigma}_t^2s_\theta(z_t,t)}{\alpha_t}$. 
This gives us the {\bf Forward Universal Guidance update}:

\begin{equation}
    \hat{s}_\theta(z_t,t) = s_\theta(z_t,t)-\nabla_{z_t} \ell\left(C,f\left(\dfrac{z_t+{\sigma}_t^2s_\theta(z_t,t)}{\alpha_t}\right)\right).
    \label{eq:forward}
\end{equation}
Here, $f$ is a user-defined differentiable function that encodes task-specific properties of the material system, such as the volume per atom, atom coordination number, bond distance, or other relevant structural or chemical descriptors. 
This differentiability ensures that the gradient $\nabla_{z_t}\ell$ in the guidance update can be computed effectively. However, as detailed in~\cite{bansal2023universal}, the Forward Universal Guidance alone does not give reasonable empirical results.
Therefore, the {\bf Backward Universal Guidance} was introduced, in order to strengthen the constraint:
\begin{equation}
    \hat{s}_\theta(z_t,t) = s_\theta(z_t,t)-\dfrac{\alpha_t}{\sigma_t^2}\nabla_{z_0} \ell\left(C,f\left(\hat{z}_{0|t}\right)\right).
    \label{eq:backward}
\end{equation}

Eq.~(\ref{eq:backward}) is similar to the training-free guidance \cite{training-free-guidance}. 
We optimize an arbitrary small perturbation $\Delta$ on the material   that best enforces $C$:
\begin{align*}
    \Delta z_0 &= \arg\min_{\Delta}\ell(C,f(\hat{z}_{0|t}+\Delta)) \\
    \hat{z}_{0|t} + \Delta z_0 &= \dfrac{z_t+{\sigma}_t^2 \left(s_\theta(z_t,t)+\dfrac{\alpha_t}{\sigma_t^2}\Delta z_0\right)}{\alpha_t}.
\end{align*}

In this perspective, one step of Backward Universal Guidance, shown as eq.~\eqref{eq:backward} is equivalent to performing one step of gradient descent with a constant step size equal to $1$ in the above mentioned optimization problem (starting from $\Delta = 0$).

Although in the original version of the classifier-free guidance~\cite{bansal2023universal},
only the Variance Preserving (VP) case was considered,
here we would like to take the best from two worlds and integrate both, VP and Variance Exploding (VE), so that we consider the general case.

Finally, the last ingredient of the Universal Guidance framework is the {\bf self-recurrence} loop. 
This loop consists of re-noising the sample according to the forward corruption process right after the denoising step. 
This is repeated $k$ times. 
The final estimation of $z_{t-1}$ is then chosen to continue the diffusion process. 
This allows a better exploration of the manifold of noisy samples at time $t$. 
All these elements result in the {\bf Universal Guidance algorithm} drafted as Algorithm~\ref{alg}, which is based on the approaches introduced in~\cite{training-free-guidance} and \cite{bansal2023universal}. The coefficients $g_s,k_s$ in Algorithm~\ref{alg} are {\it guidance strengths}
that allow for more flexibility in the guidance. 
Their utility is detailed in Section~\ref{a:guid-str}.

\begin{algorithm}
\caption{Universal Guidance}
\label{alg}

\KwIn{\vspace{-10pt}
\begin{align*}
    N: & \text{ number of denoising steps} &  n_r:& \text{ number of self-recurrence steps} \\
    n_b: & \text{ number of backward guidance steps} & s_{\theta}:& \text{ score network} \\
    a,b: & \text{ parameters of the dffusion process} & f: & \text{ guidance function} \\
    \ell: & \text{ loss function} & C: & \text{ condition for the guidance}
\end{align*}
}

\medskip

\For{$i = N$ \KwTo $1$}{
  $z'_i \leftarrow z_i$\;

  \begin{SelfRecBox}
  
  \For{$j = 1$ \KwTo $n_r$}{
    $s \leftarrow s_\theta(z'_i,t_i)$\;
    $z_{0|i} \leftarrow \dfrac{z_i'+{\sigma}_{t_i}^2s}{\alpha_{t_i}}$\;

    $s \leftarrow s
      - g_s(t_i)\cdot
      \nabla_{z'_i}
      \ell\bigl(C,f(z_{0|i})\bigr)$
      \tcp*[r]{forward guidance}

    \begin{BackGuidBox}
    
    \For{$k = 1$ \KwTo $n_b$}{
      $z_{0|i} \leftarrow \dfrac{z_i'+{\sigma}_{t_i}^2s}{\alpha_{t_i}}$\;
      $s \leftarrow s
      - k_s(t_i)\cdot
      \frac{\alpha_{t_i}}{\sigma_{t_i}^2}
      \nabla_{z_{0|i}}
      \ell\bigl(C,f(z_{0|i})\bigr)$
      \tcp*[r]{backward guidance}
    }
    \end{BackGuidBox}

    $z_{i-1} \sim \N\left(\dfrac{z_i'+b_{t_i}^2s}{a_{t_i}} ,\dfrac{{\sigma}_{t_{i-1}}^2}{{\sigma}_{t_i}^2}b_{t_i}^2\right)$\tcp*[r]{sampling}
    $z'_i \sim \N\left(a_{t_i}z_{i-1} , b_{t_i}^2\right)$
    \tcp*[r]{forward corruption}
  }
  \end{SelfRecBox}
}
\KwRet{$z_0$}
\end{algorithm}

We derive a more formal analysis of the approach, beyond the analysis originally performed by~\cite{bansal2023universal}.
By defining $(z_t^i)_{i=0}^k$ and $(z_{t-1}^i)_{i=0}^k$ and by using both forward and backward diffusion processes, we update equations as follows:

\begin{align*}
    z_{t-1}^i &= \frac{1}{a_t}\left(z_t^i+b_t^2 s_\theta(z_t^i)\right) + \dfrac{\sigma_{t-1}}{\sigma_t}b_t \cdot\varepsilon_t^i \\
z_{t}^{i+1} &= a_t\cdot z_{t-1}^i+ b_t \cdot\epsilon_t^{i+1} 
\end{align*}
where the noises $\epsilon_t^i, \varepsilon_t^i\sim \N(0,I)$ are mutually independent. Overall, the new reverse update takes the following form:
\begin{align*} 
z_{t-1}^k = \dfrac{1}{a_t}\left(z_t^i+b_t^2\sum_{i=0}^k s_\theta(z_t^i)\right) + \dfrac{\sigma_{t-1}}{\sigma_t}b_t \sum_{i=0}^k\varepsilon_t^i + \dfrac{b_t}{a_t} \sum_{i=1}^{k}\epsilon_t^{i}.
\end{align*}

Therefore, by renaming $\displaystyle\sum_{i=0}^k\varepsilon_t^i=\sqrt{k+1}\cdot\tilde{\varepsilon}_t$ and $\displaystyle\sum_{i=1}^k\epsilon_t^i=\sqrt{k}\cdot\tilde{\epsilon}_t$, these updates can be rewritten:
\begin{align*} 
z_{t-1}^k = \dfrac{1}{a_t}\left(z_t^0+\left(\sqrt{k+1}b_t\right)^2\frac{1}{k+1}\displaystyle\sum_{i=0}^k s_\theta(z_t^i)\right) + \dfrac{\sigma_{t-1}}{\sigma_t}\left(\sqrt{k+1}b_t\right) \tilde{\varepsilon}_t + \dfrac{1}{a_t}\sqrt{\dfrac{k}{k+1}} \left(\sqrt{k+1}b_t\right)\tilde{\epsilon}_t.
\end{align*}

This formulation highlights that if $k$ increases, $\frac{1}{k+1}\displaystyle\sum_{i=0}^k s_\theta(z_t^i)$ gets closer to the average score in the neighborhood of the noisy sample $z_t$, while the variance, becoming proportional to $\left(\sqrt{k+1}\cdot b_t\right)$, increases.
Therefore, a trade-off between efficient average score approximation and small variance is needed.

\subsubsection{Guidance strength}
\label{a:guid-str}
The guidance strength coefficients $g_s,k_s$ are used to control precisely the strength of the guidance during the diffusion process. Overall, three different cases are possible:
(i) increasing functions, (ii) decreasing functions, (iii) constant functions.
For example, in case where the estimator $\hat{z}_{0|t}$ is quite noisy in the first steps, we can fix $g_s(t)=k_s(t)=0$ for $t/T>0.9$. 
As explained in~\cite{training-free-guidance}, each of these cases is adapted to specific conditions. 
To reduce the computational complexity,
it is also possible to guide only at certain timesteps, for example: $g_s(t)=k_s(t)=0$ if $t\not\equiv 0~[M]$ where $M$ is the desired periodicity of the guidance. 
This decreases the guidance strength but improves the running time of the algorithm by reducing the number of operations, thereby accelerating the overall generation.

\subsubsection{Gradient normalization}

We faced a numerical issue when using off-the-shelf loss functions: the gradient of the guidance function and its derivative were not of the same order as the score.
This can lead to misguidance, as in classical gradient descent. 
To address this, especially with potentially new losses added afterward without control over their units, we implemented a slight modification of the guidance score updates by normalizing the gradient.
Our method is inspired by the normalized gradient descent method as explained in~\cite{Zhao2021stoc_grad_desc}:
\begin{align*}
    g_s(t) =  \dfrac{\gamma_t}{\| \nabla \ell(C,z_t)\|};\quad k_s(t) =  \dfrac{\kappa_t}{\| \nabla \ell(C,z_t)\|}
\end{align*}

In this modification, the coefficients $\gamma_t, \kappa_t$ play a central role: they control the guidance strength. Usually they are kept constant, but it is possible to vary them to obtain broader control over the sampling process.
More generally, Spider-SFO \cite{spider} adds an upper bound for the coefficient:
$$g_s(t) =  \min \left(\dfrac{\gamma_t}{\| \nabla \ell(C,z_t)\|}, \eta_t\right)$$

All these choices are made to simplify the theoretical smoothness condition needed for the loss function.
However, we use constant values for $\gamma_t, \kappa_t$ and select them empirically, since the above mentioned conditions are difficult to guarantee.
Finding theoretical guarantees would require an additional study.  

\subsection{Fine-tuning-free guidance for score-based stochastic differential equations generative models}

In this last methodological part, we present the algorithm that we designed for our fine-tuning-free guided generative model for material generation, Algorithm~\ref{alg2}. The diffusion model used by MatterGen is a stochastic differential equations (SDE) diffusion model. Overall, the SDE allows for more flexibility in the diffusion process, since the time is continuous rather than discrete. 
It leverages a stochastic differential equation score-based model to transform the discrete equation into a continuous process following \cite{songscore}. 
Details about the induced change in the equations are available in SI~2.2.1.
Predictor-corrector sampling (detailed in SI~2.3) is considered as a pure sampling method to be directly applied inside the universal guidance framework. 
The predictor used here is the standard sampling method for $z_{t-1}$ knowing $z_t$.
We use Algorithm~\ref{alg2} for our numerical experiments described in the following section. 

\begin{algorithm}[h!]
\caption{Universal Guidance: Predictor-Corrector}
\label{alg2}
\KwIn{\vspace{-10pt}
\begin{align*}
    N: & \text{ number of denoising steps} &  n_r:& \text{ number of self-recurrence steps} \\
    n_b: & \text{ number of backward guidance steps} & s_{\theta}:& \text{ score network} \\
    a,b: & \text{ parameters of the dffusion process} & f: & \text{ guidance function} \\
    \ell: & \text{ loss function} & C: & \text{ condition for the guidance}
\end{align*}
}

\medskip

$z_N \sim q_T$\;

\For{$i = N$ \KwTo $1$}{
  $z'_i \leftarrow z_i$\;
  \begin{SelfRecBox}
  \For{$\_ = 1$ \KwTo $n_r$}{
    $s \leftarrow s_\theta(z'_i,t_i)$\;
    $z_{0|i} \leftarrow \dfrac{z_i'+{\sigma}_{t_i}^2s}{\alpha_{t_i}}$\;

    $s \leftarrow s_\theta(z'_i,t_i)
      - g_s(t_i)\cdot
      \nabla_{z'_i}
      \ell\bigl(C,f(z_{0|i})\bigr)$
      \tcp*[r]{forward guidance}

    \begin{BackGuidBox}
    \For{$\_ = 1$ \KwTo $n_b$}{
      $z_{0|i} \leftarrow \dfrac{z_i'+{\sigma}_{t_i}^2s}{\alpha_{t_i}}$\;
      $s \leftarrow s
        - k_s(t_i)\cdot
        \frac{\alpha_{t_i}}{\sigma_{t_i}^2}
        \nabla_{z_{0|i}}
        \ell\bigl(C,f(z_{0|i})\bigr)$
      \tcp*[r]{backward guidance}
    }
    \end{BackGuidBox}
    $z_{i-1} \sim \N\left(\dfrac{z_i'+b_{t_i}^2s}{a_{t_i}} ,\dfrac{{\sigma}_{t_{i-1}}^2}{{\sigma}_{t_i}^2}b_{t_i}^2\right)$ \tcp*[r]{predictor}
    
    \begin{CorrectorBox}
    \For{$\_ = 1$ \KwTo $n_c$}{
        $\varepsilon \sim \mathcal{N}(0,I_d)$\;
        $s \leftarrow s_\theta(z_{i-1},t_{i-1})$\;
        $\lambda \leftarrow 2\alpha_{i-1}
            \left(r\frac{\|\varepsilon\|_2}{\|s\|_2}\right)^2$\;
        $z_{i-1} \leftarrow z_{i-1}
            + \lambda s
            + \sqrt{2\lambda}\varepsilon$\;
    }
    \end{CorrectorBox}
    $z'_i \sim \N\left(a_{t_i}z_{i-1} , b_{t_i}^2\right)$
    \tcp*[r]{forward corruption}
  }
  \end{SelfRecBox}
}
\KwRet{$z_0$}
\end{algorithm}

\newpage
\section{Case studies of crystal structures generated under constraints}

Having introduced the general guidance procedure above, we now turn to its main application in this work, namely inorganic crystal structure generation. 
We show how physically motivated constraints can be incorporated into diffusion sampling in order to control the generation of crystallographic cells. 
Although the discovery of new materials is not the primary objective here, we demonstrate that guidance can shift the statistical distributions of generated structures relative to non-guided sampling of the foundation model, while the generated structures are still checked for chemical consistency in the postprocessing and analysis steps described below.

We illustrate the methodology on several solid-state chemical systems taken as case studies. 
In these examples, structure generation is guided by differentiable losses associated with targeted geometric descriptors. 
A chemical system here simply refers to the set of crystalline materials formed from a given combination of chemical elements, and the corresponding guidance loss is defined according to the imposed constraints~$C$.

The effect of guidance is assessed by comparing guided and non-guided batches containing several hundred to several thousand generated structures.
Both protocols use the same MatterGen \texttt{chemical\_system} checkpoint, trained on crystal lattices with at most 20 atoms per unit cell and energies below 0.1~eV/atom above the convex hull~\cite{mattergen_checkpoints}.
The non-guided reference follows standard MatterGen sampling, whereas our guided protocol modifies the denoising trajectory through the imposed constraint loss.
Batch sizes vary across systems and guidance settings because the corresponding runs have different computational costs and were carried out under practical resource constraints. 
In our implementation, guided sampling is approximately 4.6 times more computationally expensive per generated structure than standard non-guided sampling. 
Therefore, at a fixed generation cost, guidance increases the expected number of raw constraint-satisfying structures only when the raw enrichment factor relative to non-guided sampling exceeds this cost ratio. 
Descriptor enrichment is evaluated using normalized hit fractions and statistical tests that explicitly account for the corresponding sample sizes, while the final summary table also reports the absolute number of retained targeted candidates.

Our MatterGen-based guidance implementation is available at \url{https://github.com/link-lab3629/scout-matter}; additional technical details are provided in SI~4.
For each case study, the guidance strengths $k$ and $g$ were chosen by testing a few candidate values so as to preserve chemically reasonable structures while still producing a measurable effect on the targeted descriptors.
This led us to retain relatively low values rather than enforcing the guidance constraints too strictly.
All guided runs used $N=1000$ denoising steps, $n_r=3$ self-recurrence steps (applied for $t<0.9\,T$), $n_b=2$ backward-guidance steps, $n_c=1$ corrector step inside the self-recurrence loop (Algorithm~\ref{alg2}), a Huber Loss with $\delta = 1.5$ and per-structure gradient normalization, whereas the non-guided reference uses standard MatterGen sampling without self-recurrence.

After sampling, generated data are analyzed along two complementary directions. 
First, in order to assess the direct effect of guidance on the sampling procedure itself, descriptor distributions for the guided and non-guided generated structures are computed from the raw generated data, i.e.\ before any postprocessing. 
These distributions are represented by histograms or kernel-smoothed curves and compared with the corresponding reference distributions. 
To quantify this raw descriptor-level enrichment, we use a one-sided $p$-value from a two-proportion $z$-test comparing the fractions of generated structures that satisfy the constraint in the guided and non-guided samples.
Smaller $p$-values indicate stronger raw-stage enrichment of constraint-satisfying structures in the guided sample; values below 0.05 are used here as a conventional threshold for statistical significance.

Second, both the guided and non-guided structure sets are processed through the same postprocessing pipeline, described in detail in SI~5. 
Among other steps, this pipeline includes symmetrization of nearly symmetric but slightly distorted structures, deduplication, and the removal of structures that are chemically unreasonable from the outset. 
The latter screening is performed using interatomic-distance criteria and, more importantly, by estimating the height above the energy convex hull defined by a combined dataset from the Alexandria~\cite{alexandria}, OQMD~\cite{oqmd}, and Materials Project~\cite{materials_project} databases. 
In the following, we refer to these known structures as the ``references''. 
They provide a consistent comparison set for the generated distributions, but should not be interpreted as an exact reconstruction of the MatterGen training set, since the latter involves atomic-number and energy cutoffs.

Energies are evaluated with the GRACE model, using the ``GRACE-2L-OMAT-large-ft-AM'' checkpoint~\cite{grace}, which proved to be of accuracy comparable to DFT for the systems considered here. 
We note that energy estimates are performed without crystal relaxation in order to directly assess the quality of generation and to preserve user-imposed geometric constraints, such as the target volume in our first case study. 
As an example for the Cu--Si--P subsets, the energy difference between GRACE and DFT mean absolute error ranges from 12.6 to 13.8\,meV/atom for static calculations and from 12.0 to 12.6\,meV/atom after relaxation. 
The DFT computational details, results and parity plots are provided in SI~6.
The postprocessed data are then used to analyze, by means of Pareto fronts, the relationship between constraint satisfaction and physical plausibility, with the latter approximated by the height above the convex hull (see the first example below for definitions and details).

The objective functions used in the case studies below are either the per-atom volume or local coordination descriptors. 
We denote the former as $v_\mathrm{a}=|\det L|/N_\mathrm{at}$, where $L$ is the matrix of lattice vectors and $N_\mathrm{at}$ the number of atoms in the cell. 
For coordination-based constraints, $\mathrm{CN}(A{-}B)$ denotes the average number of neighboring $B$ atoms around atoms of type $A$ within the smooth coordination descriptor used for guidance. 
Technical details of the proposed guidance functions for materials generation, including environment-dependent losses and equality constraints, are provided in SI~4. 
There, we also show that multiple guidance objectives can be combined by treating $C$ constraints as a multidimensional objective. 
More precisely, given losses $(\ell_i)_{i=1}^d$ associated with properties computed by functions $f=(f_i)_{i=1}^d$, we derive the final scalar loss as detailed in SI~4.
The framework itself is not limited to geometric descriptors: in principle, any objective differentiable with respect to atomic coordinates can be used.
As an example, we also implemented an energy-based guidance loss (SI~4.2); in practice, it performed comparably to the non-guided baseline, consistent with the fact that the MatterGen checkpoint is already trained to sample low-energy structures.

The following case studies range from unary to quaternary systems, progressively illustrating the scope of the proposed guidance strategy and how increasingly complex forms of structural control can be achieved.

\subsection{Case study 1: High-density boron} 

In the first example we explore the dense high-pressure allotropes of boron, a system known for its structural complexity and diversity~\cite{oganov_high-pressure_2011}.
Our strategy consists in enforcing a single geometric constraint: a low value of the per-atom volume, $f(z_t) = v_\mathrm{a}$.
Such low per-atom volumes are characteristic of dense boron allotropes.
For reference, the per-atom volumes of stable boron allotropes are 7.229~\AA$^3$ for $\alpha$-B, 7.730~\AA$^3$ for $\beta$-B, and for the densest known boron phase, orthorhombic $\gamma$-B, 6.982~\AA$^3$~\cite{gamma-boron}.
The chosen target, $C = v_{\mathrm{a}0} = 7.0$~\AA$^3$ per atom, therefore corresponds closely to $\gamma$-B.
Hence, the guidance loss function is $\ell(C,f(z_t)) = \|v_\mathrm{a} - v_{\mathrm{a}0}\|_1$.

\begin{figure}
    \centering
    \includegraphics[width=0.9\linewidth]{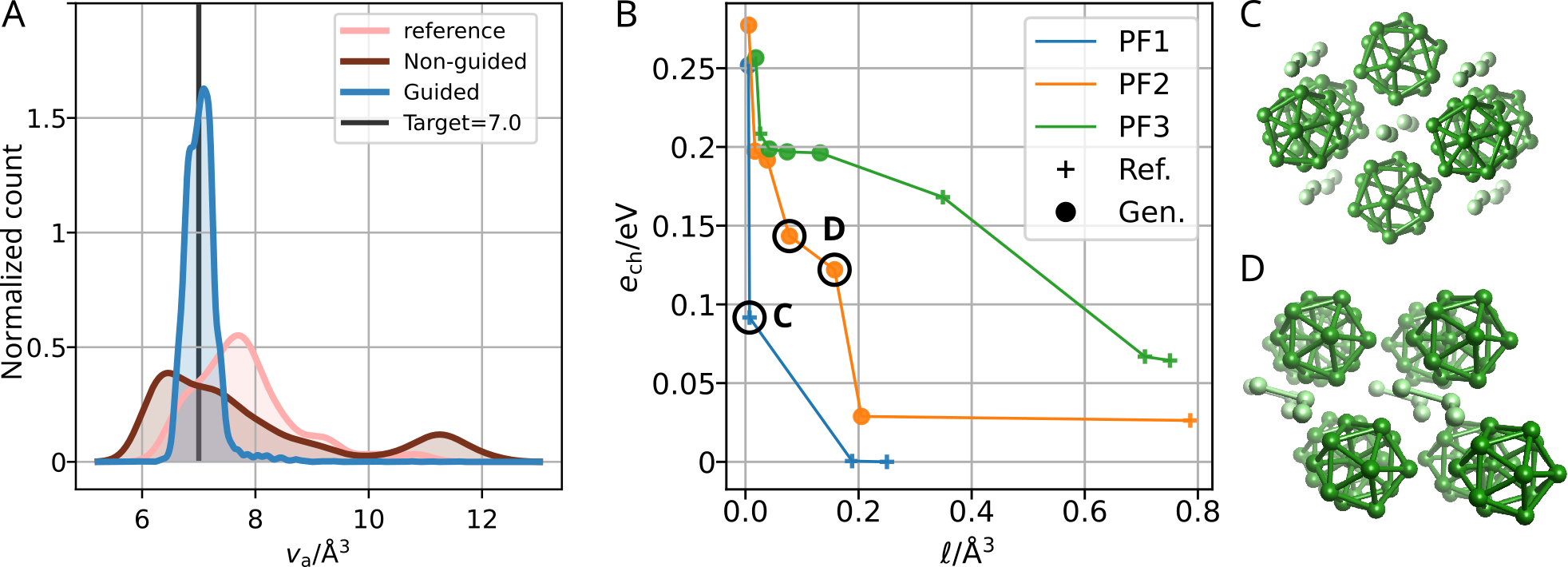}
    \caption{A. Distributions of per-atom volumes in the reference dataset and in the datasets generated by non-guided and guided sampling with $k=g=0.5$ toward a target value of 7.0~\AA$^3$. B. First three Pareto fronts (PF1-3) in the $\ell$ vs.\,$e_\mathrm{ch}$ coordinates for the mixture of the generated (Gen.) and reference (Ref.) datasets for high-density boron allotropes. C. Structure of $\gamma$-boron. D. Structure representing two nearly identical best generated lattices from the second Pareto front.}
    \label{fig:B-combined}
\end{figure}

The $\gamma$-B allotrope presents 28 atoms in the unit cell, which is relatively few compared to the most common boron phases (\textit{cf.} $\beta$-rhombohedral boron with 105 atoms), yet it is the most recent phase to be discovered~\cite{oganov_high-pressure_2011}. 
Because the 28-atom unit cell exceeds MatterGen's 20-atom limit, $\gamma$-B itself is excluded from the checkpoint by construction. 
The size mismatch is not extreme, so this boron case tests whether guidance can approach a nearby high-density structural regime rather than simply reproduce an in-domain training example. 
In Fig.~\ref{fig:B-combined}, we present the results summary of guided generation. 
Fig.~\ref{fig:B-combined}.A, shows the distributions of the per-atom volume for the generated structures, together with the corresponding distributions for the 44 reference allotropes available in the databases. 
The curves show Gaussian-kernel-smoothed distributions, with smearing selected by Silverman's rule of thumb\cite{silverman}.

The non-guided generation roughly reproduces the reference distribution with the mean value around 7.8~\AA$^3$ and a small peak around 11~\AA$^3$ corresponding to 2D-borophene phase, for which the database entries are periodic lattices with vacuum layers separating the sheets, hence high volume.

The main result is the pronounced redistribution under guidance: about 60~\% of the guided samples fall within $\pm 0.25$~\AA$^3$ of the target value of 7.0~\AA$^3$, compared to $\sim 20$~\% in the non-guided case. 
This difference is not only visually apparent in Fig.~\ref{fig:B-combined}.
A but also statistically significant: a two-proportion $z$-test comparing the fractions of near-target structures in the guided and non-guided samples yields $p < 10^{-12}$. 
These results provide strong quantitative evidence that guidance effectively concentrates sampling around the desired volume.

To put these near-target samples in a materials-science context, we complement structural-feature matching with a basic thermodynamic validity check by estimating the per-atom energy above the convex hull envelope formed by the reference structures. 
This value, $e_\mathrm{ch}$, compares a given crystal with the most stable competing mixture of reference entries of the same overall empirical formula.
Non-positive $e_\mathrm{ch}$ indicates thermodynamic stability. 
For elemental boron, $e_\mathrm{ch}$ is simply the per-atom energy relative to the lowest-energy known allotrope.

To jointly consider target proximity and low $e_\mathrm{ch}$, we rank structures in the two-dimensional space defined by the guidance loss and $e_\mathrm{ch}$. 
In this ranking, a structure is better than another one if it has lower or equal loss and lower or equal $e_\mathrm{ch}$, with a strict improvement in at least one of the two quantities. 
The first Pareto front contains the structures for which no better alternative is found in the population. 
Subsequent fronts are then constructed iteratively: after removing the structures already assigned to fronts $1,\ldots,i-1$, the $i$-th front contains the best remaining structures according to the same criterion.

The Pareto fronts for generated and reference data combined are shown in Fig. \ref{fig:B-combined}.B. 
The guided model successfully identified two very similar dense structures consisting of B$_{12}$ clusters with interstitial B$_{3}$ units (Fig. \ref{fig:B-combined}.D), which closely resemble the $\gamma$-boron allotrope (Fig. \ref{fig:B-combined}.C). 
This outcome underscores the efficacy of our guidance mechanism in exploring crystal structures with densities near the target. 
While the identified structure is not an exact match to $\gamma$-boron, its proximity highlights the potential of our approach for refining and guiding materials prediction.

\subsection{Case study 2: Boron coordination in the Fe--Nd--B system}

The Fe--Nd--based compounds are widely studied for their properties as permanent magnets, particularly for data storage applications.
To enhance the localization of $d$-electrons of the transition metal (Fe), one effective strategy is to introduce neutral elements such as boron~\cite{Sagawa84}. 
In fact, boron serves a structural role by increasing the Fe--Fe distance, which enhances the ferromagnetic moment and coercivity. 
The precise manner in which boron modulates Fe--Fe exchange interactions is critical to achieving superior magnetic hardness.
Several stable compounds are known in this system, such as the boron-rich phases Nd(FeB)$_4$ and Nd$_5$(FeB$_3$)$_2$, where B is coordinated by 3 to 6 Fe atoms depending on the site. 
Among the known iron-rich phases, and thus of particular interest for magnetism, is Nd$_2$Fe$_{14}$B, where Fe atoms form continuous puckered hexagonal nets linked by short Fe--Fe bonds,
separated by planes containing both Nd and B atoms~\cite{Nd2Fe14B, herbst_1984}. 
The compound adopts a tetragonal structure (space group $P4_2/mnm$), comprising 68 atoms per unit cell.  
Since this is too large to be reproduced by MatterGen generation, we address the boron coordination sphere as a proxy of the overall crystal structure:
in Nd$_2$Fe$_{14}$B, B atoms occupy the centers of trigonal prisms [BFe$_6$], with Fe--B bond distances of 2.09--2.14\,\AA. 
These trigonal prisms are fundamental structural units that link the Fe layers perpendicular to the planes containing Nd, B, and some additional Fe atoms. 
In this case study, our aim is to generate structures that maintain these [BFe$_6$] building blocks, by defining the geometric constraint for the generation. 
This approach favors the essential structural motifs that contribute to the exceptional magnetic hardness and energy density of the parent Nd$_2$Fe$_{14}$B compound, while focusing the generation task on a local coordination feature that can be evaluated within the current model constraints.

\begin{figure}
    \centering
    \includegraphics{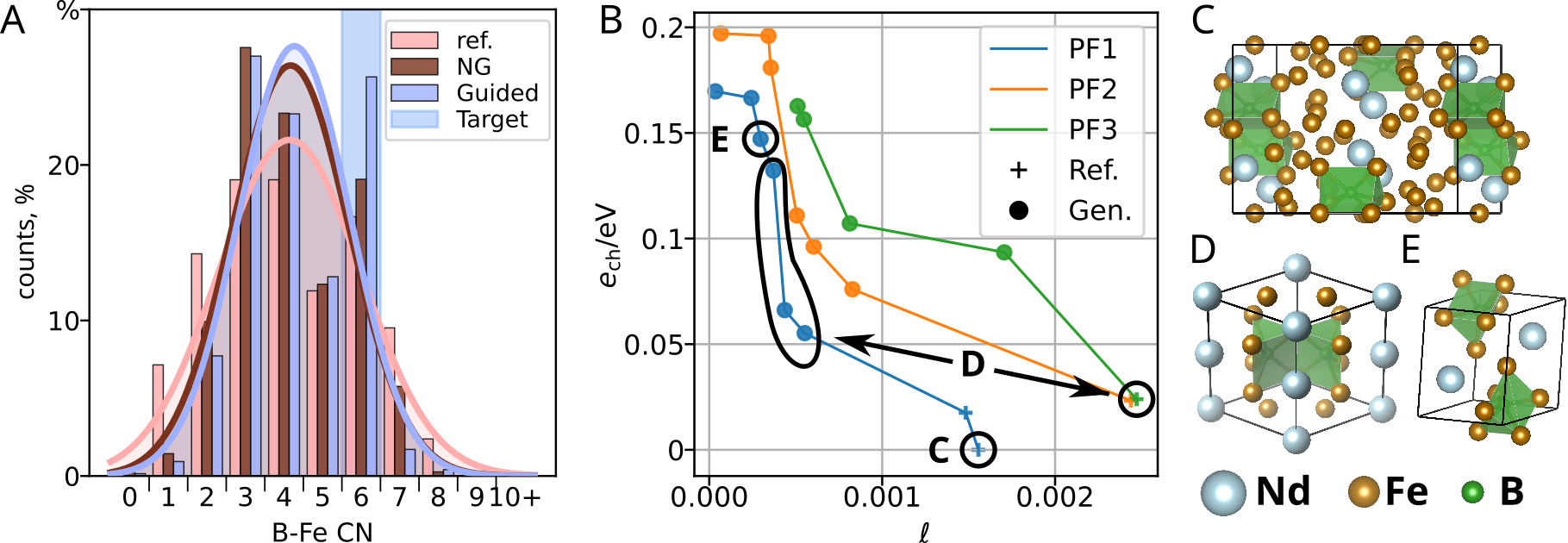}
    \caption{Fe--Nd--B system. A. Counts of structures by B--Fe coordination number (CN) in the reference dataset and in non-guided and guided generations, the latter with $k=g=0.5$ and target CN $=6$. Each distribution is supplied with a gaussian curve with matching mean and variance. B. First three Pareto fronts PF1--3 for the guided+reference dataset. C. Reference Nd$_2$Fe$_{14}$B structure. D. Compact NdNi$_4$B-type motif: the reference appears on PF3, with slightly distorted generated versions on PF1. E. A distinct low-loss generated candidate at $\sim$0.15~eV/atom above the reference hull.}
    \label{fig:Fe-Nd-B}
\end{figure}

Figure~\ref{fig:Fe-Nd-B} summarizes the sampling of B--Fe coordination, following the same analysis protocol as in the elemental boron example discussed above.
All three distributions in Fig.~\ref{fig:Fe-Nd-B}A are broad and have nearly identical mean and variance, as illustrated by gaussian curves matching the true mean and variance which we provide for a rough first-glance analysis. 
All three exhibit pronounced maxima at coordination numbers 3, 4 and 6. 
Guidance, however, produces a clear enrichment of the targeted six-fold environment, demonstrating that the sampling can be steered toward a prescribed B--Fe coordination while retaining a realistic spread of local motifs. 
The effect of the guidance is statistically significant, with a $p$-value of $2.03\times10^{-10}$. 
After the full filter chain, the guided sample retains 18 unique, original, metastable, and target-satisfying candidates. 
Structural inspection of the low-loss, low-energy region of the Pareto plot shows what is recovered.
The most stable Fe-rich Nd$_2$Fe$_{14}$B phase shown in Fig.~\ref{fig:Fe-Nd-B}C contains the ideal magnetic [BFe$_6$] environment but is outside the present range of numbers of atoms in MatterGen.
Within this range, however, the guided run recovers NdFe$_4$B with the NdNi$_4$B-type motif~\cite{NdNi4B} with planar hexagonal network of edge-sharing BFe$_6$ prisms. Several generated candidates are slightly distorted versions of the reference, which appears on the third Pareto front (Fig.~\ref{fig:Fe-Nd-B}D).
The same guided run also produces an unseen low-loss candidate that remains reasonably metastable, at about 0.15~eV/atom above the reference hull (Fig.~\ref{fig:Fe-Nd-B}E).
Thus, within the current size cap, guided generation reproduces the reference structure and also proposes a new metastable candidate respecting the targeted BFe$_6$ local constraint.

\subsection{Case study 3: Co--O environment in Li--Co--O}

The ternary Li--Co--O system is well-known for its prototypical compound, lithium cobalt oxide LiCoO$_2$, a cornerstone cathode material in lithium-ion batteries~\cite{MIZUSHIMA1980}. 
Its layered structure enables reversible lithium intercalation, a critical feature for energy storage applications. 
LiCoO$_2$ exists in two primary crystalline forms: a high-temperature (HT) hexagonal phase and a low-temperature (LT) cubic spinel-related phase, with the HT-LiCoO$_2$ rhombohedral structure being the preferred choice for battery applications~\cite{ANTOLINI2004}. 
For this latter compound, Li$^+$ cations are intercalated between sheets of edge-sharing CoO$_6$ octahedra, while cobalt in its +3 oxidation state balances the charge of the two O$^{2-}$ anions.
The $d^6$ low-spin electron configuration of Co$^{3+}$ favors octahedral coordination, splitting $d$-orbitals to stabilize the structure through crystal field effects. 

Our aim in this case study is to deliberately enrich coordination environments that are not dominant in the training database.
By doing so, we seek to demonstrate that reduced coordination numbers, such as fourfold Co--O environments, can be enforced, enabling targeted in silico exploration of candidate crystalline structures.

\begin{figure}
    \centering
    \includegraphics{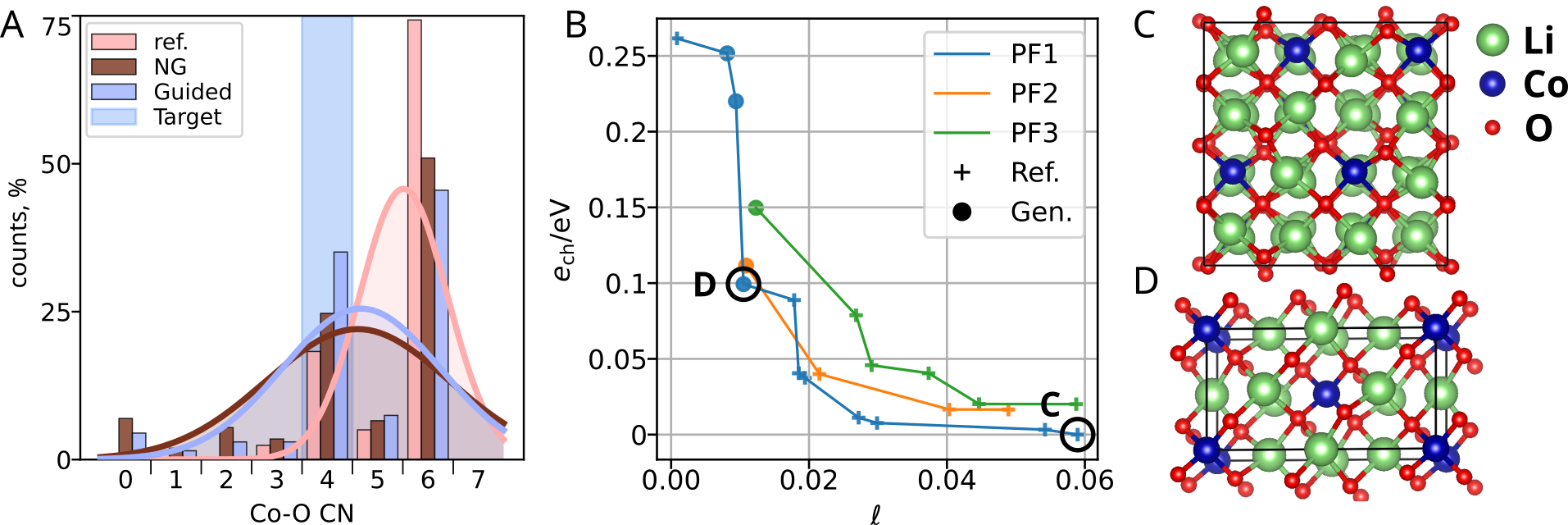}
    \caption{Li--Co--O system: A. Counts of structures by average Co--O coordination number (CN) in reference dataset and datasets obtained from non-guided (NG) and guided generations ($k=1.0$, $g=0.01$, target = 4). B. First three Pareto fronts for guided+reference dataset, C. reference HT-Li$_{5}$CoO$_{4}$. 
    D. Li$_3$CoO$_{4}$ recovered from guided generation.}
    \label{fig:Li-Co-O}
\end{figure}

The results of guidance toward 4-fold Co--O coordination are shown in Fig.~\ref{fig:Li-Co-O}. 
This motif, commonly associated with Co$^{2+}$, appears as the second-most populated coordination in both the reference dataset and the non-guided samples (Fig.~\ref{fig:Li-Co-O}A). 
Applying guidance substantially enhances this feature, and the effect is statistically significant ($p = 0.021$).
The lowest-energy reference structure with near-zero Co--O coordination loss, HT-Li$_5$CoO$_4$ (Fig.~\ref{fig:Li-Co-O}C), is a spinel-like compound related to the Li$_5$FeO$_4$ prototype\cite{li5feo4} and contains 80 atoms per cell, placing it outside the current MatterGen cell-size limit.
The guided run instead yields a much smaller constraint-satisfying tetragonal Li$_3$CoO$_4$ candidate with 8 atoms per primitive cell (Fig.~\ref{fig:Li-Co-O}B and D), which is nevertheless low in energy, at $\sim$0.10~eV/atom above the reference hull.




\subsection{Case study 4: Cu--P environment in ternary Cu--Si--P} 
The Cu--Si--P chemical system is of particular interest due to its ternary compounds, which exhibit semiconductor properties and may serve as high-capacity anode materials in lithium-ion and redox flow batteries, thanks to low volume expansion compared to silicon-based anodes~\cite{CuSiP-1}.
Compounds like CuSi$_2$P$_3$ and CuSi$_3$P$_4$ adopt distorted zinc-blende structures~\cite{wang_composition_2010}, 
making them relevant for studies in optoelectronics and energy band gap engineering. 
While experimental syntheses of CuSi$_2$P$_3$ report a disordered arrangement of Cu and Si cations with largely covalent bonding in a tetrahedrally coordinated phosphorus environment~\cite{wang_composition_2010,CuSiP-2}, DFT calculations from the Materials Project suggest a nearly stable ordered phase with CuP$_4$ and SiP$_4$ tetrahedra~\cite{CuSiP-3}.
A similar chemical environment is observed in Cu$_4$SiP$_8$, where Cu atoms form Cu--Cu pairs which are octahedrally coordinated to six P~\cite{CuSiP-5}.
When combined with alkaline earth elements, other coordination environments appear, for example in BaCuSi$_2$P$_3$, where  copper is bonded to 3 phosphorus and 2 silicon atoms. 
The Cu atoms are in a near trigonal planar environment with the 3~P atoms~\cite{CuSiP-4}. 

\begin{figure}
    \centering
    \includegraphics{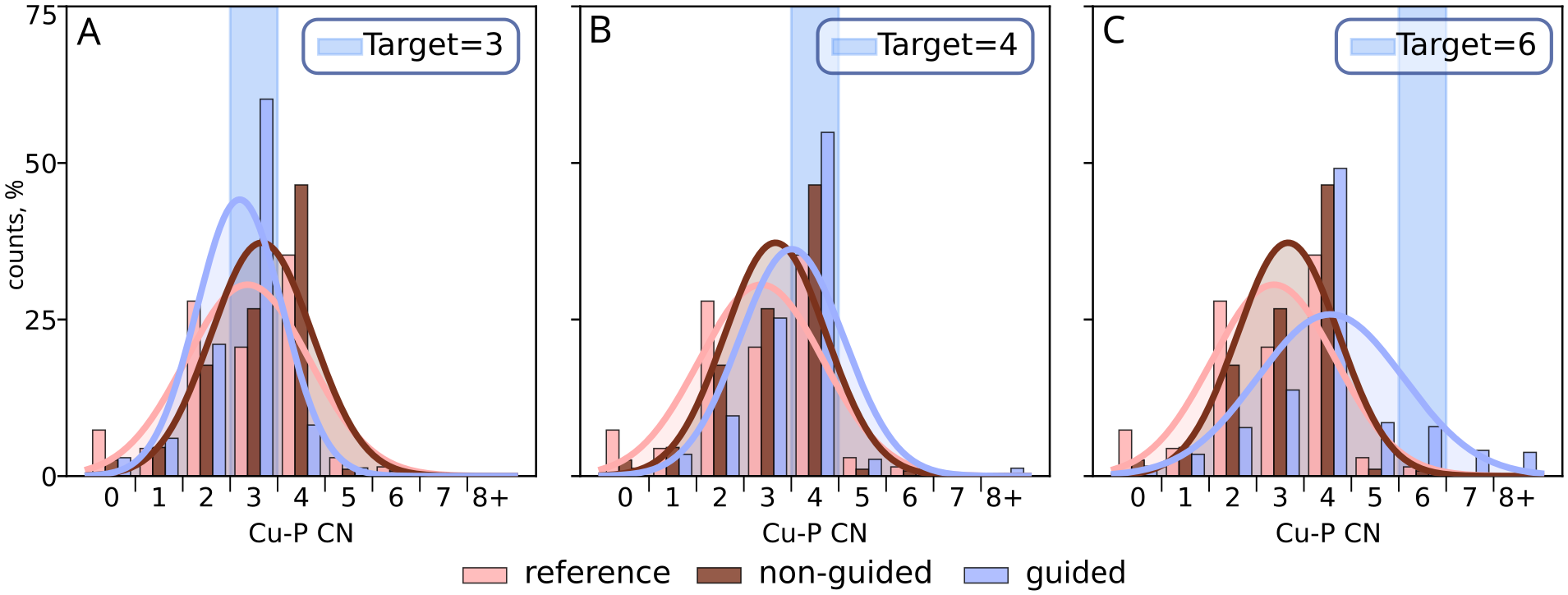}
    \caption{Counts of structures by Cu--P coordination numbers (CN), for three target phosphorus environments of Cu atoms in Cu--Si--P system. Guidance was carried out with $k=g=1.0$.}
    \label{fig:Cu-Si-P}
\end{figure}

Here we used the Cu--Si--P system as a controlled testbed to quantify how strongly our guidance mechanism can reshape local-environment statistics, independently of whether the targeted motif is thermodynamically plausible, by imposing three distinct Cu--P coordination targets (3, 4, and 6 neighbors). 
In the reference data, fourfold Cu--P coordination is prevalent, whereas sixfold coordination is essentially absent; targeting the latter therefore provides a stringent stress test of the steering capability. 
As shown in Fig.~\ref{fig:Cu-Si-P}, guidance produces clear, systematic shifts in the coordination-number peaks relative to non-guided sampling ($p<10^{-12}$, $3.73\times10^{-3}$, and $2.65\times10^{-11}$ for the Cu--P targets 3, 4, and 6, respectively), increasing the population of structures that satisfy the requested environment even in the extreme sixfold case. 
While such a coordination is chemically atypical and likely energetically unfavorable in this composition space, demonstrating that the generator responds predictably to a local structural constraint is valuable: it establishes guidance as a statistical control knob that can be used to generate counterfactual and deliberately imbalanced structure sets. 
These controlled distributions can, in turn, support downstream machine-learning workflows,
for example, augmenting training data for environment-conditioned generators, constructing hard negative examples for coordination classifiers, or improving the robustness and calibration of surrogate property models and machine-learning interatomic potentials by exposing them to rare (or deliberately perturbed) local motifs.

\subsection{Case study 5: Multiple constraints in Cu--Si--P--Ca quaternary system}

Following our previous case study on the ternary Cu--Si--P system, we extend the investigation to the quaternary Cu--Si--P--Ca chemical space, for which no quaternary entries are present in our selected reference databases.
We chose Ca to move beyond the reported ternary family into a less charted quaternary setting while remaining within the broader compositional family of alkaline-earth-containing Cu--Si--P phosphides, such as layered BaCuSi$_2Pn_3$ ($Pn=\mathrm{P},\mathrm{As}$)\cite{CuSiP-4}. 
This provides a demanding setting in which to test whether motif-directed generation remains effective once the search is lifted into a broader compositional space.

Within this quaternary system, we target a Cu--Cu dimer stabilized in a distorted octahedral phosphorus environment, as found in CuP$_2$ and Cu$_4$SiP$_8$.
The interest of this motif lies in the fact that it couples a direct metal--metal contact to a specific phosphorus coordination scaffold, so that successful generation requires recovering not only Cu--Cu pairing but also the local anion environment that supports it.

\begin{figure}
    \centering
    \includegraphics[width=0.8\linewidth]{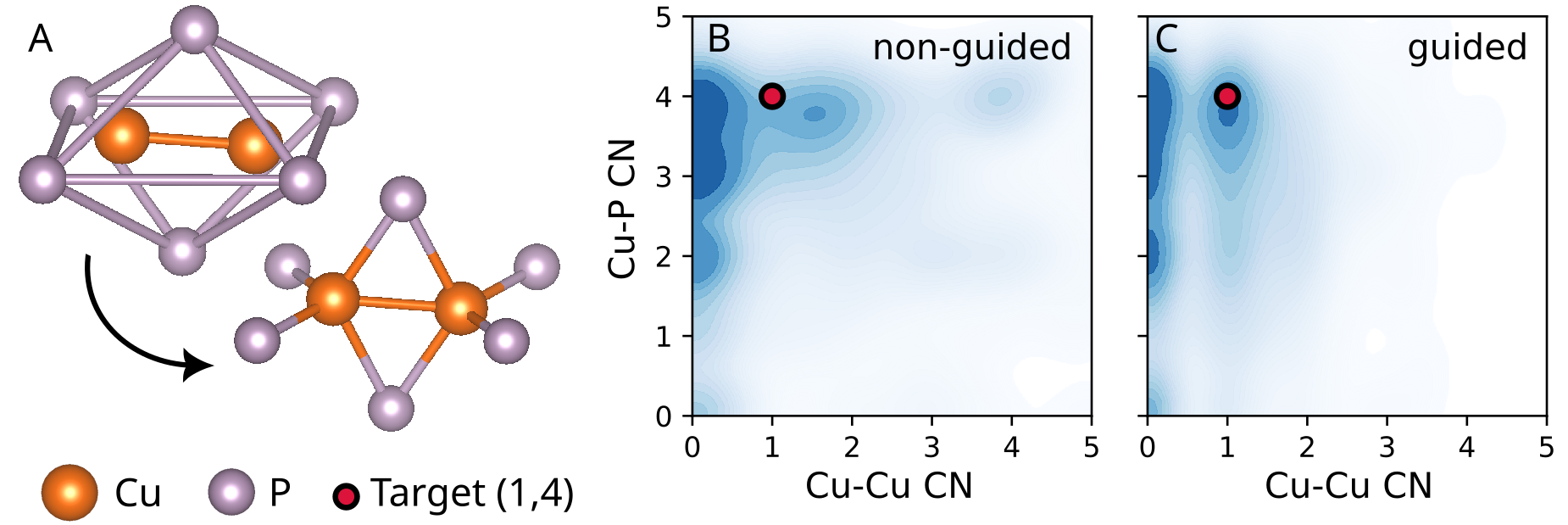}
    \caption{A. Target structural pattern of a Cu--Cu dimer within a distorted P octahedron, equivalently represented as two edge-sharing CuP$_4$ tetrahedra. B. Kernel-smoothed distribution of average Cu--Cu and Cu--P coordinations for non-guided sampling of the Cu--Si--P--Ca system. C. Same for guided generation with constraints CN(Cu--Cu)~=~1, CN(Cu--P)~=~4 (red dot) ($k=g=0.8$).}
    \label{fig:multi}
\end{figure}

In the previous examples, guidance was aimed directly at isolated coordination environments. 
Here, by contrast, the objective is a composite local motif rather than a single coordination feature. 
We show that such a motif can nevertheless be targeted using multi-objective guidance built from simple coordination-number constraints. 
Specifically, we impose $\mathrm{CN}(\mathrm{Cu{-}Cu}) = 1$ and $\mathrm{CN}(\mathrm{Cu{-}P}) = 4$ for each Cu atom. 
In this representation, the target Cu$_2$P$_6$ motif is decomposed into two edge-sharing CuP$_4$ tetrahedra (Fig.~\ref{fig:multi}A). 
The point is therefore not that the constraints themselves are more elaborate, but that their combination is sufficient to specify a structurally nontrivial local arrangement.

The results of sampling are illustrated in Fig.~\ref{fig:multi}B,C, where we compare the 2D kernel density estimates obtained without and with guidance. 
In the unguided case (Fig.~\ref{fig:multi}B), the distribution is dominated by high-density regions corresponding to structures without direct Cu--Cu bonding, while the target appears only as a weak saddle-like feature. 
With multi-objective guidance (Fig.~\ref{fig:multi}C), the dominant peaks are largely retained, but a pronounced additional local maximum emerges at the target, showing that the guidance selectively enriches the desired motif rather than simply redistributing the entire ensemble. This is supported by a $p$-value below $10^{-12}$.

Importantly, the appearance of a local maximum at the target in Fig.~\ref{fig:multi}C demonstrates only that the combined coordination constraints are satisfied more frequently under guidance; it does not, by itself, establish recovery of the intended Cu$_2$P$_6$ motif, since the same values of $\mathrm{CN}(\mathrm{Cu{-}Cu})$ and $\mathrm{CN}(\mathrm{Cu{-}P})$ may in principle also arise from alternative local geometries.
For this reason, structural inspection is essential.

\begin{figure}
    \centering
    \includegraphics[width=12cm]{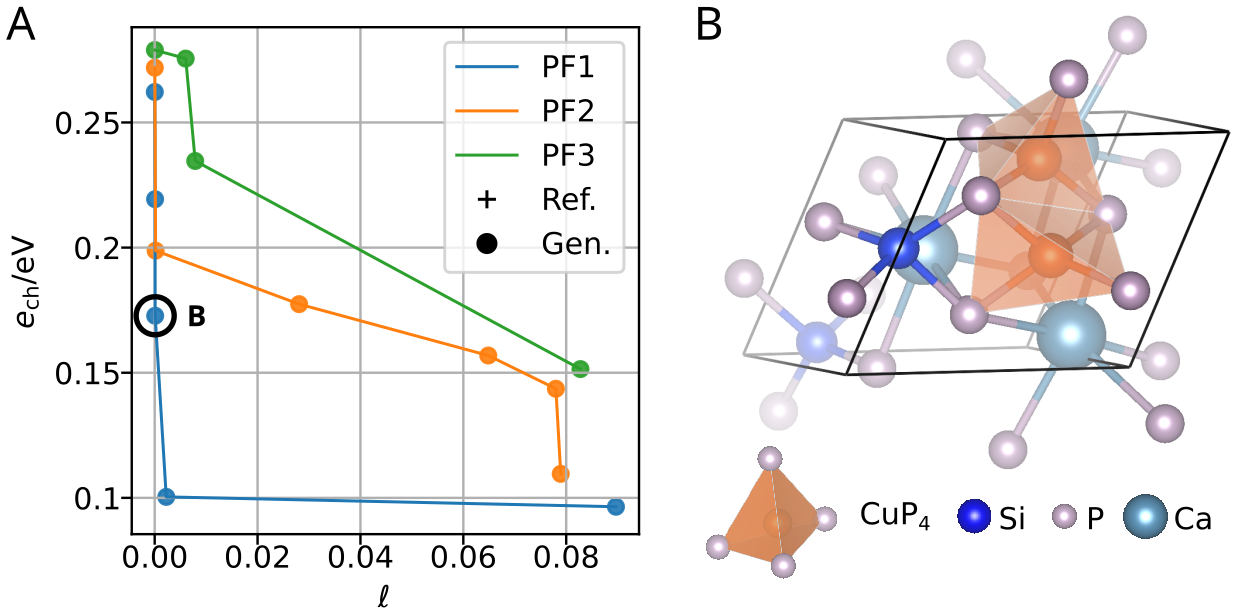}
    \caption{A. First three Pareto fronts in energy above convex hull versus loss coordinates. B. Representative zero-loss Cu$_2$Si$_2$P$_5$Ca$_3$ structure found from guided generation.}
    \label{fig:multi_pareto}
\end{figure}

The Pareto analysis in Fig.~\ref{fig:multi_pareto}A identifies a set of low-loss, low-energy candidates.
The representative zero-loss structure Cu$_2$Si$_2$P$_5$Ca$_3$ shown in Fig.~\ref{fig:multi_pareto}B, lying 0.175~eV/atom above the convex hull, confirms that the guided search can recover not only the formal coordination counts but also the intended arrangement of a Cu--Cu dimer in a distorted octahedral phosphorus environment, realized as two edge-sharing CuP$_4$ tetrahedra rather than, for example, a stacked square-planar-like alternative.
The Cu--Cu bond length is 2.27~\AA, which, even without relaxation, is close to reported values of 2.51~\AA{} in Cu$_4$SiP$_8$ and 2.48~\AA{} in CuP$_2$~\cite{CuSiP-5}.
Taken together, these results show that combined local descriptors can do more than enrich the prescribed coordination counts: within the structural prior learned by MatterGen, they can guide sampling toward low-energy motifs compatible with those counts.

\subsection{Summary of results and discussion}

The main result of this work is that fine-tuning-free guidance can steer a pre-trained crystal diffusion model toward user-defined structural descriptors, such as local coordination motifs or volume-related targets.
To make this comparison explicit, Table~\ref{tab:stagewise_retention_nonrelaxed_focus} uses $R$ for the set of raw generated structures and $C$ for the set of structures satisfying the target constraint.
The raw targeted fraction $|C|/|R|$, together with the corresponding one-sided enrichment $p$-value, shows that the guided run enriches the targeted descriptor relative to the matched non-guided baseline in all showcase systems.
The size of this steering effect is target-dependent: it is very large for dense boron and for several Cu--P coordination targets, modest for B--Fe and Co--O coordination, and strong at the raw stage for the combined Cu--P/Cu--Cu motif.
Thus, at the descriptor level, the guidance loss does what it is designed to do: it reallocates probability mass in the generated distribution without retraining the foundation model.

The key question is how many structures generated under guidance remain scientifically reasonable after post-processing, and how many of those still satisfy the target constraints.
Table~\ref{tab:stagewise_retention_nonrelaxed_focus} therefore also follows each generated sample through deduplication, the metastability-window filter, database matching, and the final target-constraint filter. Here, $U$ denotes the set retained after deduplication (Unique); $S$, the set of structures predicted within 0.2~eV/atom of the reference hull (metaStable); and $O$, the set of structures not matched to the selected reference databases (Original). The non-guided rows are evaluated using exactly the same definition of $C$ as the corresponding guided runs.

We separate two quantities that answer related but non-equivalent questions. 
First, we define the conditional constrained-retention factor
\[
\mathrm{SCOUT}=\frac{|U\cap S\cap O\cap C|}{|U\cap S\cap O|}
\]
which asks whether the intended motif is retained among candidates already passing the $U$, $S$, and $O$ filters.
We use the name SCOUT as a shorthand for stable, constrained, original, unique target recovery; the metric itself is defined by the ratio above and is applied identically to guided and non-guided samples.
The raw-normalized counterpart
\[
Y_\mathrm{SCOUT}=\frac{|U\cap S\cap O\cap C|}{|R|}
\]
instead asks how many useful targeted candidates are obtained per raw generated structure.

Both views are needed: SCOUT measures conditional selectivity after chemical filtering, whereas $Y_\mathrm{SCOUT}$ measures the final yield of targeted candidates from the full generation run.
We also report the absolute count $|U\cap S\cap O\cap C|$, since small normalized yields can correspond to only a few actual structures. 
The table is evaluated on the full generated samples without further structural optimization; the reported $p$-value is the raw-stage enrichment test shown once per guided/non-guided pair.

\begin{table}[htbp]
\centering
\scriptsize
\setlength{\tabcolsep}{1.4pt}
\renewcommand{\arraystretch}{0.90}
\caption{Stage-wise retention for the showcase case studies. The table reports raw constraint enrichment and stage-wise conditional retention through deduplication, metastability, originality, and the final constraint filter; $|U\cap S\cap O\cap C|$ gives the corresponding absolute count, and $Y_\mathrm{SCOUT}=|U\cap S\cap O\cap C|/|R|$ is the final SCOUT yield.}
\label{tab:stagewise_retention_nonrelaxed_focus}
\begin{tabular}{llcrrccccccc}
\toprule
System & Target & Set & $|R|$ & $\frac{|C|}{|R|}$ & $p$ & $\frac{|U|}{|R|}$ & $\frac{|U \cap S|}{|U|}$ & $\frac{|U \cap S \cap O|}{|U \cap S|}$ & $|U\cap S\cap O\cap C|$ & $\mathrm{SCOUT}$ & $Y_\mathrm{SCOUT}$ \\
\midrule
B & $V/N=7$ & non-guided & 2000 & 0.127 & $<10^{-12}$ & 0.661 & 0.005 & 0.833 & 4 & 0.800 & 0.002 \\
 &  & guided & 6312 & 0.600 &  & 0.691 & 0.002 & 0.889 & 7 & 0.875 & 0.001 \\
\midrule
B--Fe--Nd & B--Fe $=6$ & non-guided & 4300 & 0.155 & $2.03\times10^{-10}$ & 0.284 & 0.389 & 0.992 & 18 & 0.038 & 0.004 \\
 &  & guided & 3000 & 0.212 &  & 0.354 & 0.196 & 0.981 & 18 & 0.088 & 0.006 \\
\midrule
Co--Li--O & Co--O $=4$ & non-guided & 1200 & 0.052 & 0.021 & 0.787 & 0.675 & 0.995 & 39 & 0.061 & 0.033 \\
 &  & guided & 600 & 0.077 &  & 0.793 & 0.651 & 0.990 & 20 & 0.065 & 0.033 \\
\midrule
Cu--Si--P & Cu--P $=3$ & non-guided & 2500 & 0.138 & $<10^{-12}$ & 0.724 & 0.541 & 0.990 & 145 & 0.150 & 0.058 \\
 &  & guided & 708 & 0.431 &  & 0.983 & 0.103 & 0.986 & 27 & 0.380 & 0.038 \\
\midrule
Cu--Si--P & Cu--P $=4$ & non-guided & 2500 & 0.404 & $3.73\times10^{-3}$ & 0.724 & 0.541 & 0.990 & 298 & 0.308 & 0.119 \\
 &  & guided & 716 & 0.459 &  & 0.888 & 0.258 & 0.994 & 117 & 0.718 & 0.163 \\
\midrule
Cu--Si--P & Cu--P $=6$ & non-guided & 2500 & 0.016 & $2.65\times10^{-11}$ & 0.724 & 0.541 & 0.990 & 7 & 0.007 & 0.003 \\
 &  & guided & 706 & 0.059 &  & 0.820 & 0.297 & 0.994 & 0 & 0.000 & 0.000 \\
\midrule
Cu--Si--P--Ca & \makecell[l]{Cu--P $=4$\\Cu--Cu $=1$} & non-guided & 2200 & 0.010 & $<10^{-12}$ & 0.742 & 0.555 & 0.986 & 13 & 0.015 & 0.006 \\
 &  & guided & 1132 & 0.072 &  & 0.718 & 0.226 & 0.978 & 4 & 0.022 & 0.004 \\
\bottomrule
\end{tabular}
\end{table}

Reading SCOUT and $Y_\mathrm{SCOUT}$ together gives a compact view of the showcase cases. Cu--P $=4$ is the clearest success: SCOUT increases from 0.308 to 0.718, and the final SCOUT yield increases from 0.119 to 0.163. B--Fe $=6$ is also a positive, although more moderate, case: SCOUT increases from 0.038 to 0.088, the final SCOUT yield increases from 0.004 to 0.006, and the Pareto analysis shows recovery of compact small-cell analogues of the intended [BFe$_6$] motif. Co--O $=4$ is nearly neutral after filtering, with a small SCOUT increase from 0.061 to 0.065 and essentially unchanged $Y_\mathrm{SCOUT}$. For dense boron, Cu--P $=3$, and the combined Cu--P/Cu--Cu target, guidance improves conditional SCOUT but not the final yield. Cu--P $=6$ remains the clearest final-filter failure: despite raw-stage enrichment, the guided run produces no final $U\cap S\cap O\cap C$ candidate.

Because the energetic screening in Table~1 is performed on unrelaxed generated geometries, one possible concern is that the guidance gradients may displace structures from nearby local minima and thereby artificially reduce their retention at the metastability-filtering stage. 
Although the MatterGen checkpoint provides a prior favoring low-energy crystal structures, this prior does not guarantee that guided samples remain close to local minima.
To assess this effect while limiting the computational cost, we performed DFT relaxations on representative structures randomly selected from one non-guided subset and two guided Cu--Si--P subsets.  
Computational details and the DFT relaxation-energy distributions are provided in SI~6. 
For the investigated guided subsets, the median relaxation energy, defined as ($\Delta E_{\mathrm{relax}}=E_{\mathrm{static}}-E_{\mathrm{relaxed}}$), ranges from 2.09 to 2.49~meV/atom, and approximately 96\% of the structures in each subset exhibit an energy change below 10\,meV/atom. 
The corresponding relaxation-energy distributions are comparable to that of the non-guided baseline. 
These small energy changes show that most structures in the investigated samples are already energetically close to their DFT-relaxed configurations. 
Thus, for these Cu--Si--P targets, guided structures are not systematically farther in energy from nearby DFT local minima than the non-guided structures.
Consequently, the lower retention observed for some of these guided subsets cannot be explained solely by evaluating their energies before relaxation. 

These small energy changes show that most structures in the investigated samples are already energetically close to their DFT-relaxed configurations. Thus, for these Cu--Si--P targets, guided structures are not systematically farther in energy from nearby DFT local minima than the non-guided structures.

One possible interpretation of low final retention rates is that the guided run is not simply the non-guided distribution conditioned on $C$.
The unguided model samples from a learned joint distribution in which local descriptors, composition, geometry, and energetic plausibility are implicitly correlated.
External guidance increases the frequency of the target descriptor, but may also favor descriptor-level shortcuts: geometries that reduce the differentiable loss without preserving the broader structural context that made the same motif compatible with metastability in the unguided distribution.
This would explain why the guided run can show stronger raw enrichment in $C$ while producing fewer structures in $U\cap S\cap O\cap C$. 
Deduplication and database matching remain useful controls against collapse and rediscovery, but in the present showcase the main yield loss usually occurs at the metastability step.
The lack of a fine-tuned baseline remains a limitation of this study. 
Our results do not imply that training-free guidance would outperform a conditional model trained on abundant and representative target-specific data. 
Its intended advantage is the ability to explore rare or changing constraints for which such data are not readily available.
The central value of fine-tuning-free guidance is therefore not that it replaces crystal-structure prediction, nor that it guarantees a higher metastable--unique--original--constrained yield. 
Its value is controllability: a pre-trained crystal generator can be redirected toward interpretable structural hypotheses without retraining or constructing new conditional datasets, and in favorable cases this improves both descriptor selectivity and final candidate yield. 
The remaining challenge is to make this control less disruptive to the learned structural prior. 
This points to differentiable descriptors with more physically faithful gradients and to variable guidance-force profiles during a single sampling run, rather than a fixed post-warmup strength. 
In this role, guidance is a targeted proposal mechanism that expands how crystal foundation models can be queried, while still leaving thermodynamic evaluation and structural interpretation as essential filters.

\newpage

\section{Conclusions}

In this work, we have shown that training-free guidance can provide a pretrained crystal diffusion model with inference-time control over explicit structural objectives, including atomic volume, local coordination environments, and combinations of near-neighbor motifs. The principal result is not a uniformly improved yield of stable candidate structures, but the ability to redirect the sampling distribution toward interpretable, user-defined structural hypotheses without retraining the underlying model. Across the investigated systems, this descriptor-level control ranges from strong enrichment to more modest effects and, in some cases, does not survive the subsequent energetic and structural filters. 
These results establish training-free guidance as a flexible proposal mechanism whose usefulness depends on the target and chemical system, rather than as a general replacement for conditional fine-tuning or established crystal-structure-prediction workflows.
Training-free guidance should therefore be viewed as complementary to fine-tuning, particularly when the targeted structural regime is sparsely represented or when constraints must be modified without constructing a new training dataset.

This framework paves the way for a more directed exploration of materials space, where crystallographic cell distributions can be oriented toward user-defined local environments and densities, providing a flexible proposal tool for the assisted design and analysis of functional inorganic materials.

\section*{Data availability}
The reference crystal structures used for comparison were obtained from the Alexandria, OQMD and Materials Project databases, as cited in the manuscript.
A representative set of raw guided and non-guided generation archives used by the reproducibility workflow is distributed with \texttt{vsbtools} at \url{https://github.com/link-lab3629/vsbtools/tree/master/vsbtools/materials_dataset/Examples/raw_generations}.
The workflow regenerates the processed datasets, descriptor statistics, energy-above-hull estimates, Pareto-front data and figures in a separate output directory and records the run configuration in a machine-readable manifest.

\section*{Code availability}
The full computational workflow is publicly available in two repositories maintained by the authors' laboratory.
Crystal generation and the fine-tuning-free guidance implementation are provided in \texttt{scout-matter} (\url{https://github.com/link-lab3629/scout-matter}), including installation and command-line instructions for guided and non-guided runs.
Postprocessing, reference-database matching, energy and convex-hull analysis, deduplication, statistical analysis and figure generation are provided in \texttt{vsbtools} (\url{https://github.com/link-lab3629/vsbtools}).
A self-contained setup script, packaged example inputs and an executable end-to-end notebook are located in \texttt{vsbtools/materials\_dataset/Examples}; step-by-step and headless execution instructions are given in SI~7.

\section*{Acknowledgments}
This research is part of an extensive, multidisciplinary collaboration within the international LINK unit (CNRS-NIMS-Saint-Gobain) and the French Investments for the Future Program (PIA) as part of the France 2030 Initiative operated by the National Agency for Research (ANR) through the PEPR DIADEM priority program, within the MADNESS initiative (Artificial intelligence-assisted methodology for the discovery of new materials in molten salts, grant agreement 23-PEXD-0011).
F.d'A.-B. and N.S. would like to thank the Isaac Newton Institute for Mathematical Sciences, Cambridge, for support and hospitality during the programme ``Representing, calibrating and leveraging prediction uncertainty from statistics to machine learning'' where part of the work on this paper was undertaken. This programme was supported by EPSRC grant no EP/Z000580/1. F.d'A.-B. also received
funding from the European Union's Horizon Europe research and innovation programme under grant
agreement 101120237 (ELIAS).
DFT calculations were performed using HPC resources from GENCI--CINES (Grant A0060906175).
We warmly thank developers of VESTA, which was essential for reproducing the crystal structure drawings in this study~\cite{momma_vesta_2011}.

\section*{Author contributions}

A.d.L. developed and implemented the fine-tuning-free guidance framework, adapted the MatterGen-based sampling workflow, performed the guided-generation experiments, carried out statistical analyses, and participated in the writing and the revision of the methodological sections of the manuscript; V.B. designed the atomic environment guidance loss, developed and applied the postprocessing and validation workflow, including deduplication, reference-database comparison, energy-above-hull analysis, Pareto-front analysis, structural inspection, carried out generations and subsequent analysis for the case studies, co-wrote and prepared figures for the corresponding section; D.P. contributed to the chemical motivation, selection and interpretation of the inorganic case studies, and critical assessment of the generated structures from a solid-state chemistry perspective, and participated in manuscript writing and revision; G.L. co-conceived the project, supervised the generative-AI and materials-discovery aspects, contributed to the design of the guidance strategy and validation protocol, and participated in manuscript writing and revision; N.S. contributed to the machine-learning formulation, guidance-loss design, statistical interpretation, and critical revision of the manuscript; F. d'.A.-B. originally proposed to use guidance mechanism, contributed to its design, co-supervised the machine learning aspects of the project and revised the manuscript; J.--C.C. co-conceived and supervised the project, defined the materials-science objectives and chemical constraints, coordinated the validation strategy, contributed to the interpretation of the results, and participated in manuscript writing and revision. A.d.L. and V.B. contributed equally.

All authors discussed the results, revised the manuscript, and approved the final version.

\subsection*{Declaration of generative AI and AI-assisted technologies}

During the revision of the manuscript, the authors used AI-assisted technologies for English-language editing, including grammar, clarity, and readability, and for drafting limited portions of source code from author-defined specifications. 
All AI-assisted text was reviewed and edited by the authors. 
All AI-assisted code was manually reviewed, modified where necessary, and validated through unit tests, integration tests, and the continuous-integration workflow. 
The disclosed LLM-based tools were not used to autonomously define the scientific methodology, generate or interpret the reported results, or formulate the scientific conclusions.

\section*{Competing interests}
The authors declare no competing interests.

\newpage

\bibliography{bibliography}

\includepdf[pages=-]{article_guidance-si.pdf}
\end{document}